\documentclass[fleqn,twoside]{article}
\usepackage{espcrc2}
\usepackage{graphicx}
\usepackage{epsfig}
\title{Pion form factor with twisted mass QCD}

\author{
        Abdou M.~Abdel-Rehim\address[UofR]{Department of Physics, 
        University of Regina, Regina, SK, S4S 0A2, Canada} and
        Randy Lewis\addressmark[UofR]}
       
\begin{document}

\begin{abstract}
The pion form factor is calculated using quenched twisted mass QCD 
with $\beta=6.0$
and maximal twisting angle $\omega=\pm\frac{\pi}{2}$. 
Two pion masses and several values of momentum transfer are considered.
The momentum averaging procedure of Frezzotti and Rossi is used to reduce
lattice spacing errors, and numerical results are consistent with the
expected $O(a)$ improvement.
\vspace{1pc}
\end{abstract}
\maketitle

\section{INTRODUCTION}
Lattice QCD with a twisted mass term (tmQCD) is a computationally
efficient method for eliminating the exceptional configurations that
plague light Wilson quarks.\cite{tmQCD}
In addition, at maximal twist, many quantities are $O(a)$ improved
and others become $O(a)$ improved through a momentum averaging
procedure.\cite{FreRos}
The pion form factor provides an opportunity to explore this improvement
procedure.

With only two valence fermions and no disconnected lattice
diagrams\cite{Draper},
the pion form factor is an appealing laboratory for studies of the transition
between the perturbative and nonperturbative regimes of QCD.
Experimental measurements are available for comparison, and current experiments
at Jefferson Lab are exploring higher momentum transfers.\cite{experiment}

Recent lattice studies of the pion form factor have considered various lattice
actions, with and without $O(a)$ improvement, and the effects of $O(a)$ terms
are found to be non-negligible.\cite{lattice-results}

In this work, we present results from pion form factor calculations using tmQCD
and compare to existing lattice studies.

\section{METHOD}
Computations are performed with the $\beta=6.0$ Wilson gauge action and the
twisted mass action for a degenerate doublet of up and down
quarks with no Symanzik improvement (clover) term.\cite{tmQCD}
A twisting angle of $\pm\frac{\pi}{2}$ is obtained by setting the hopping
parameter
to its critical value, $\kappa_c=0.156911$.\cite{Jansen:2003ir} The quark mass
is then determined by the remaining parameter $\mu$ in the tmQCD action, and
results
are reported here for two options, $|\mu|=0.030$ and $0.015$, corresponding
to pion masses near $660$ and $470$ MeV respectively.
We use 100 configurations that are $16^3\times48$ with periodic boundary
conditions.
The GMRES-DR matrix inverter\cite{gmres-dr}, which deflates the
smallest eigenvalues and systematically improves them upon successive restarts
of the standard GMRES iteration, was used throughout this work.

\begin{figure}[htbp]
\includegraphics[width=75mm]{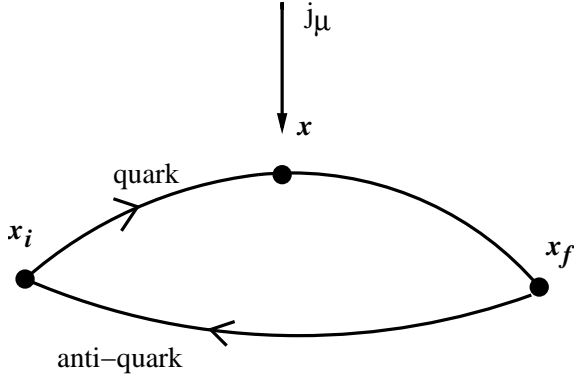}
\vspace{-12mm}
\caption{Three point correlator for the pion form factor.}
\label{3point}
\end{figure}

The pion form factor $F(Q^2)$ is defined by
\begin{equation}
\left<\pi^+(\vec{p}_f)|j_\mu(0)|\pi^+(\vec{p}_i)\right>=F(Q^2)(p_i+p_f)_\mu
\end{equation}
where $j_\mu(0)$ is a conserved vector current evaluated at the spacetime
origin, $p_i$ and $p_f$ are the initial and final pion (Euclidean) 4-momenta 
respectively, $\vec p_i$ and $\vec p_f$ are the corresponding 3-momenta,
and $Q^2=(p_f-p_i)^2$ is the 4-momentum transfer.  To compute
the above matrix element on a spacetime lattice, one can use the three point
correlator displayed in Fig. \ref{3point}.
A source with pion quantum numbers is placed at $x_i$, a sink at $x_f$,
and a vector current is inserted at $x$.
The pion form factor is extracted from a simultaneous single exponential fit
to the long time range of the two and three point correlators given by:
\[
G_{\pi\pi}(t_i,t,\vec p) = \sum_{\vec x}e^{-i(\vec x-\vec x_i)\cdot\vec p}
  \left<0|\phi(x)\phi^\dagger(x_i)|0\right>
\]
\begin{equation}
\stackrel{t \gg t_i}{\to}~~
\frac{|Z|^2}{E}
e^{-\frac{TE}{2}}
\cosh\left[\left(t-t_i-\frac{T}{2}\right)E\right],
\end{equation}
\[
\Gamma_{\pi\mu\pi}(t_i,t,t_f,\vec p_i,\vec p_f)=
\sum_{\vec x_i,\vec x_f}
e^{-i(\vec x_f-\vec x)\cdot\vec p_f}
e^{-i(\vec x-\vec x_i)\cdot\vec p_i}
\]
\[
\left<0|\phi(x_f)j_\mu(x)\phi^\dagger(x_i)|0\right>
\stackrel{t_f \gg t \gg t_i}{\longrightarrow}
\]
\begin{equation}
\frac{|Z|^2
e^{-(t-t_i)E_i-(t_f-t)E_f}}{4E_iE_f}
\left<\pi^+(\vec p_f)|j_\mu(0)|\pi^+(\vec p_i)\right>
\end{equation}
where
\begin{equation}
\left<0|\phi(x)|\pi^+(\vec{p})\right>=Ze^{ip\cdot x},
\end{equation}
$\phi(x)$ is a local interpolating field operator with $\pi^+$ quantum
numbers, $T$ is the temporal extent of the lattice and $E$, $E_i$, $E_f$
denote pion energies.
In practice, we compute one propagator from $x_i$ to $x$, and a double
propagator from $x_i$ to $x_f$ to $x$.  Three different options for the double
propagator were studied, corresponding to the local pseudoscalar operator at
$x_f$ having momentum $\vec p_f=(0,0,0)$, $(0,0,p_{\rm min})$ and
$(0,0,-p_{\rm min})$, where
$p_{\rm min} = \frac{2\pi}{L}$ (L=16).  In all cases, $x_i$ and $x_f$ are
separated by 15 time slices.  The conserved
vector current was used at $x$ and a local pseudoscalar at $x_i$.
Smeared operators have been used routinely in pion form factor
studies\cite{lattice-results}, but we chose local operators for this first
consideration of tmQCD.

\begin{table}
\caption{Pseudoscalar and vector meson masses.}
\label{ps-vec-mass-table}
\begin{tabular}{|l|l|l|}
\hline
      & $|\mu|=0.015$ & $|\mu|=0.030$ \\
\hline
$am_\pi$ & 0.238(5) & 0.331(3) \\
$am_\rho$ & 0.453(37) & 0.496(19) \\
\hline
\end{tabular}
\end{table}
\begin{figure}[htbp]
\includegraphics[width=75mm]{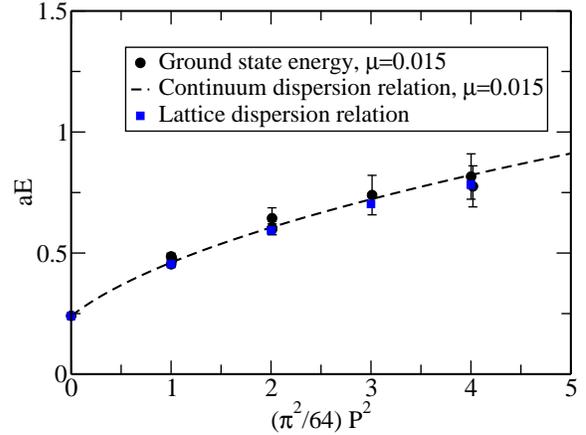}
\vspace{-12mm}
\caption{Pion dispersion relation at $|\mu|=0.015$.}
\label{ps-dispersion}
\end{figure}

\begin{figure}[htbp]
\includegraphics[width=75mm]{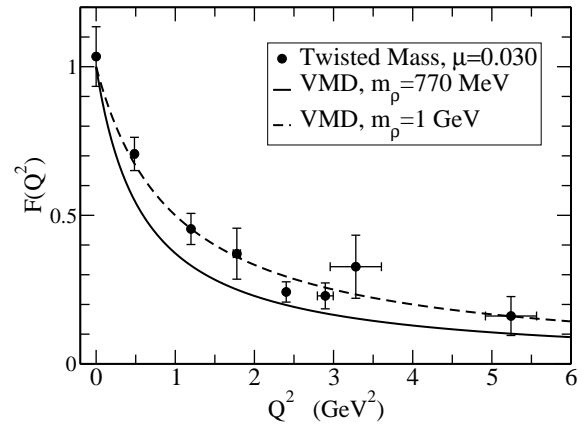}
\vspace{-12mm}
\caption{Form factor at $|\mu|=0.030$ compared to vector meson
  dominance (VMD).}
\label{FF-03}
\end{figure}

\begin{figure}[tb]
\includegraphics[width=75mm]{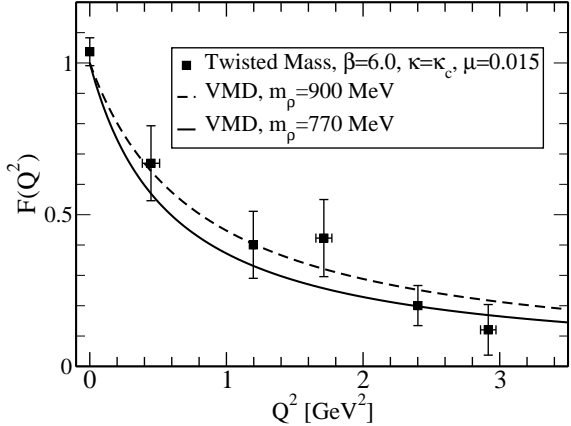}
\vspace{-12mm}
\caption{Form factor at $|\mu|=0.015$ compared to vector meson
  dominance (VMD).}
\label{FF-015}
\end{figure}

\begin{figure}[hbt]
\includegraphics[width=75mm]{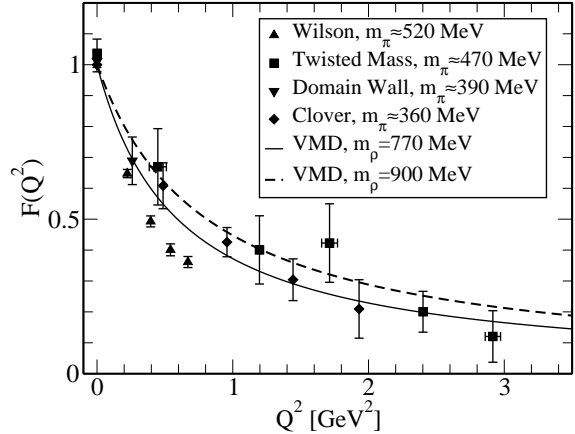}
\vspace{-12mm}
\caption{Comparing quenched twisted mass results with earlier quenched
calculations.}
\label{FF-compare}
\end{figure}
\section{RESULTS}
Analyzing the pseudoscalar-pseudoscalar and vector-vector two point
correlators, we found that a reasonable signal
could be obtained for momenta
$|\vec{p}|^2 \le 4p_{\rm min}^2$.
Figure \ref{ps-dispersion} shows the dispersion relation for the ground state
pseudoscalar meson at $|\mu|=0.015$. Table \ref{ps-vec-mass-table} lists
results for the pseudoscalar and vector meson masses in lattice units.

To extract the form factor at higher $Q^2$, nonzero sink momentum is important,
but then improvement requires momentum averaging over
positive and negative momenta.\cite{FreRos}  Figures \ref{FF-03} and
\ref{FF-015} show our results for the pion form factor as a function of $Q^2$.

A comparison with existing literature is shown in Fig. \ref{FF-compare}.
Notice that unimproved Wilson results are
systematically below experiment (which follows vector meson dominance in this
region of $Q^2$).
Decreasing $m_\pi$ decreases $F(Q^2)$ even further.
However, the tmQCD results are consistent with experiment
and with $O(a)$ improved actions.

\section*{Acknowledgements}

Helpful communications with Walter Wilcox regarding GMRES-DR are greatly
appreciated.
This work was supported in part by the Natural Sciences and Engineering Research
Council of Canada, the Canada Foundation for Innovation, the Canada Research
Chairs program, and the Government of Saskatchewan.  Some of the computing was
done with WestGrid facilities.

\end{document}